\documentclass[a4paper,12pt]{article}
\usepackage{amsfonts,amsthm,amsmath,amssymb,hyperref,youngtab}

\def\vac{|0\rangle}
\def\bra#1{\left\langle #1\right|}

\def\corr#1{\,\left\langle\, #1 \,\right\rangle\,}
\newcommand{\tr}{\operatorname{tr}}
\def\Tr{{\rm Tr\, }}

\def\id{\textrm{id}}

\def\Dim{\textrm{Dim}}

\def\r{\rho}

\def\s{\sigma}

\def\d{\partial}

\def\t{\tau}

\def\R{\mathbb{R}}
\def\Z{\mathbb{Z}}

\newcommand{\cA}{\mathcal A}
\newcommand{\cB}{\mathcal B}

\newcommand{\cG}{\mathcal G}

\newcommand{\cN}{\mathcal N}
\newcommand{\cO}{\mathcal O}

\newcommand{\be}{\begin{equation}}
\newcommand{\bea}{\begin{eqnarray}}

\newcommand{\ee}{\end{equation}}
\newcommand{\eea}{\end{eqnarray}}

\newcommand{\nn}{\nonumber}

\begin{document}
{}~
{}~
\hfill\vbox{\hbox{}\hbox{QMUL-PH-07-08} 
}\break

\vskip 3cm

\centerline{{ \bf \Large Half-BPS $SU(N)$ Correlators in $\mathcal{N}=4$ SYM   }}


\medskip

\vspace*{4.0ex}

\centerline{\large \rm 
T.W.\, Brown${}^\star$}

\vspace*{4.0ex}
\begin{center}
{\large Centre for Research in String Theory\\
Department of Physics\\
Queen Mary, University of London\\
Mile End Road\\
London E1 4NS UK\\
}
\end{center}

\vspace*{5.0ex}

\centerline{\bf Abstract} \bigskip

In this note we study half-BPS operators in $\cN = 4$ super Yang-Mills
for gauge group $SU(N)$ at finite $N$.  In particular we elaborate on
the results of hep-th/0410236, providing an exact formula for the null
basis operators algorithmically constructed there.  For gauge groups
$U(N)$ and $SU(N)$ we show that this basis is dual to the basis of
multi-trace operators with respect to the two point function.  We use
this to extend the results of hep-th/0611290 concerning factorisation
and probabilities from $U(N)$ to $SU(N)$.  We also give a construction
for a separate diagonal basis of the $SU(N)$ operators in terms of the
higher Hamiltonians of the complex matrix model reduction of this
sector.

\thispagestyle{empty}

\vfill
\noindent{\it  {${}^\star$t.w.brown@qmul.ac.uk}}
\eject

\section{Introduction}

The matching of the physics of the half-BPS sectors of $\cN=4$
super Yang-Mills in 4 dimensions and gravity in $AdS_5 \times S^5$
provides an important confirmation of the AdS/CFT
correspondence \cite{malda}.  Half-BPS operators are particularly
tractable because their quantum numbers are not renormalised and
certain extremal correlators are protected from renormalisation.  This
means that their correlators can be computed in the free theory, taken
to strong coupling and then, via AdS/CFT, compared at large $N$ with
gravity \cite{fmmr}.

In $\cN=4$ SYM half-BPS operators are built from traceless symmetric
$SO(6)$ tensor combinations of the six real scalars $X_i$, traced over
their gauge indices (the $X_i$ transform in the adjoint representation
of the gauge group).  We will be interested in the subset of those
operators built from a single complex scalar $\Phi = X_1 + iX_2$,
invariant under the remaining $SO(4)$ subgroup of the $SO(6)$
symmetry.  The conformal structure of the theory dictates the
correlator to be
\begin{equation}
  \corr{\Phi^\dagger_a(x) \Phi_b(y)} = \frac{ g_{ab}}{(x-y)^2}
\end{equation}
where $a,b$ run over the adjoint representation of the gauge group and
$g_{ab}$ is the inverse of the bilinear invariant form $g^{ab} =
\tr(T^a T^b)$.  From now on we will drop the spacetime dependence of
the correlators, because we are only interested in their
group-theoretic structure.

For the $U(N)$ gauge group the adjoint representation of the Lie
algebra consists of $N^2$ $N\times N$ hermitian matrices.  If we
consider the matrix indices of $\Phi^i_j = \Phi_a (T^a)^i_j$, where
$T^a$ is an element of the adjoint representation of the Lie algebra
of $U(N)$, we find
\begin{equation}
  \langle\; \Phi^{\dagger i}_{\phantom{\dagger}j} \Phi^k_l \;\rangle  =
  g_{ab}(T^a)^i_j (T^b)^k_l=  \delta^i_l \delta^k_j
\end{equation}
The space of gauge-invariant chiral primary operators of a particular
dimension in this $SO(4)$-invariant sector is made of products of
traces (`multi-traces') of $\Phi$.  The number of fields $\Phi$ in the
operator gives both the scaling dimension and the R-charge of the
operator, which is a typical BPS saturation condition.  In \cite{cjr}
the authors showed that linear combinations of the multi-trace
operators called Schur polynomials diagonalise this two point function
at finite $N$.

For dimension $k \ll N$ mixing between the trace operators is
suppressed, so we map $\tr(\Phi^k)$ to a graviton with angular
momentum $k$ around the the $X_1 - X_2$ plane of the sphere of $AdS_5
\times S^5$. If $k \sim N$ mixing between trace operators is no longer
suppressed so we must look instead to the diagonal Schur polynomials
for the appropriate objects on the gravity side.  These correspond to
D3 branes spinning in the geometry, called giant gravitons
\cite{mst,cjr}.  As a complex matrix model the eigenvalues correspond
to fermions in a harmonic potential \cite{cjr, Berenstein:2004kk} and
there is an exact map between the fermion distribution and the
corresponding gravity solution with $\R \times SO(4) \times SO(4)$
symmetry \cite{llm}.

For the $SU(N)$ gauge group elements of the Lie algebra are in
addition traceless and the correlator receives a correction
\begin{equation}
  \langle\;  \Psi^{\dagger i}_{\phantom{\dagger}j} \Psi^k_l   \;\rangle  =
  \delta^i_l \delta^k_j - \frac{1}{N}\delta^i_j \delta^k_l
\end{equation}
Although at large $N$ mixing between trace operators is still
suppressed, at finite $N$ this correction to the correlator
complicates the combinatorics significantly.  The Schur polynomials
are no longer diagonal.  In \cite{deMelloKoch} a basis of the $SU(N)$
gauge-invariant operators called the null basis was found, which,
while not diagonal, still has extremely nice properties, including a
simple correlator.  We will clarify the r\^ole of this basis here.

$U(N)$ is equivalent to $SU(N) \times U(1)$ up to a $\Z_N$
identification.  In the gauge theory the $U(1)$ vector multiplet is
free, so the corresponding $AdS$ field must decouple from all other
fields living in the bulk, since gravity couples to everything.  The
field is a singleton field that lives at the boundary of $AdS$,
corresponding to the centre of mass of the D3 branes
\cite{singletons}.

In Section \ref{unsum} we will summarise the known $U(N)$ results and
introduce the dual basis and its properties.  Section \ref{sunsum}
covers the corresponding $SU(N)$ picture, which is expanded upon in
Section \ref{sundetails} with detailed proofs.  Section \ref{factor}
extends the factorisation results of \cite{copto} from $U(N)$ to
$SU(N)$ and Section \ref{diag} describes the diagonalisation in terms
of the higher Hamiltonians of the complex matrix model.  There are
some useful identities in Section \ref{notation}.

\section{$U(N)$ summary}\label{unsum}

For $U(N)$ theories the correlator for the complex scalar is
\begin{equation}
  \langle\;  \Phi^{\dagger i}_{\phantom{\dagger}j} \Phi^k_l   \;\rangle  =
  \delta^i_l \delta^k_j
  \label{uninnerproduct}
\end{equation}
We have three bases for the gauge invariant multi-trace
polynomials of $\Phi$. 
\begin{enumerate}
\item The \textbf{trace basis}, of products of traces of $\Phi$ such
  as $\tr(\Phi\Phi)\tr(\Phi)$, is the obvious gauge-invariant basis.
  These multi-traces at level $n$ are in one-to-one correspondence
  with the $p(n)$ conjugacy classes\footnote{Conjugacy classes of
    $S_n$ encode the different cycle structures of permutations.} of
  the permutation group $S_n$ where $p(n)$ is the number of partitions
  of $n$.  Define a set of elements $\{\s_I\}$ in the permutation
  group $S_n$ where each $\s_I$ is an element of a different conjugacy
  class of $S_n$.  All the possible multi-trace operators of dimension
  $n$ are given by the $p(n)$ operators
  \begin{equation}
    \tr (\s_I \Phi) = \sum_{j_1, j_2, \dots j_n} \Phi^{j_1}_{j_{\s_I(1)}}
    \Phi^{j_2}_{j_{\s_I(2)}} \cdots  \Phi^{j_n}_{j_{\s_I(n)}}
  \end{equation}
  For example an element $\s_I$ of $S_5$ made up of two 1-cycles and a
  3-cycle, such as $\s_I = (1)(3)(245)$, gives an element of the trace
  basis $\tr(\s_I \Phi) = \tr(\Phi)\tr(\Phi) \tr(\Phi^3)$.

\item The \textbf{Schur polynomial basis} is defined as a sum of these
  trace operators over the elements $\s$ of $S_n$, weighted by the
  characters of $\s$ in the representation $R$ of $S_n$
  \begin{equation}
    \chi_R(\Phi) = \frac{1}{n!}\sum_{\s \in S_n}  \chi_R(\s) \tr (\s \Phi)
  \end{equation}
  The representations $R$ of $S_n$ can be labelled by Young diagrams
  with $n$ boxes, which also correspond to partitions of $n$. Thus
  there are $p(n)$ Schur polynomials of degree $n$.  $R$ also
  corresponds to a representation of $U(N)$.\footnote{For a unitary
    matrix $U$ the character of $U$ in the representation $R$ is given
    by $\chi_R(U)$ defined by this formula.  That $R$ is a
    representation of both $S_n$ and $U(N)$ is a consequence of the
    fact that $U(N)$ and $S_n$ have a commuting action on $V^{\otimes
      n}$, where $V$ is the fundamental representation of $U(N)$.}

  The correlation function of two Schur polynomials is diagonal for
  any value of $N$ \cite{cjr}
  \begin{equation}
    \corr{  \chi_R (\Phi^\dagger) \chi_S(\Phi)  }  = \delta_{RS}
    f_R
    \label{diagonalschur}
  \end{equation}
  $f_R$ is computed by
  \begin{equation}
    f_R =
    \frac{n! \;\Dim_R }{d_R} =  \prod_{i,j} (N-i+j)
    \label{fR}
  \end{equation}
  where $\Dim_R$ is the dimension of the $U(N)$ representation $R$ and
  $d_R$ is the dimension of the symmetric group $S_n$
  representation $R$. In the product expression we sum over the boxes of
  the Young diagram for $R$, $i$ labelling the rows and $j$ the columns.
  
  We can invert the relation between traces and Schur polynomials
  using the identities in Section \ref{notation}
  \begin{equation}
    \tr(\s_I \Phi) = \sum_{R(n)}  \chi_R(\s_I) \chi_R(\Phi)
  \end{equation}
  where we sum over representations $R$ of $S_n$ with Young diagrams
  of $n$ boxes.  This gives us a compact formula for the correlation
  function of two elements of the trace basis
  \begin{equation}
    \corr{  \tr (\s_I \Phi^\dagger)\tr (\s_J \Phi) }  = \sum_R f_R
    \chi_R(\s_I) \chi_R(\s_J)
    \label{untracecorrelator}
  \end{equation}

\item Define the $p(n)$ elements of the \textbf{dual basis} by
  \begin{equation}
    \xi(\s_I,\Phi):= \frac{|[\s_I]|}{n!} \sum_{R(n)} \frac{1}{f_R}\chi_R(\s_I) \chi_R(\Phi)
  \end{equation}
  where $|[\s_I]|$ is the size of the conjugacy class of $\s_I$.  Note
  that $\xi(\s_I,\Phi)$ is constant on the conjugacy class of $\s_I$.

  This basis is useful because it is dual to the trace basis using the
  inner product defined in \eqref{uninnerproduct}, i.e.
  \begin{equation}
    \corr{  \xi(\s_I, \Phi^\dagger)  \tr (\s_J \Phi)  }  = \delta_{IJ}
    \label{UNbasisisdual}
  \end{equation}
  We can show this using the diagonality of the Schur polynomials
  \eqref{diagonalschur} and the identity \eqref{eq:sumRorthogonal2} in
  Section \ref{notation}
  \begin{align}
    \corr{  \xi(\s_I, \Phi^\dagger)  \tr (\s_J \Phi)  }  & =
    \frac{|[\s_I]|}{n!} \sum_{R(n)} \frac{1}{f_R}\chi_R(\s_I)
    \sum_{S(n)} \chi_S(\s_J) \corr{\chi_R(\Phi^\dagger) \chi_S(\Phi)}
    \nn \\
     & = \frac{|[\s_I]|}{n!} \sum_{R(n)}\chi_R(\s_I)
     \chi_R(\s_J) \nn \\
     & = \delta_{IJ}
  \end{align}
  The correlation function of two elements of the dual basis is given
  by
  \begin{equation}
    \corr{  \xi(\s_I, \Phi^\dagger)\xi(\s_J, \Phi) }  =
    \frac{|[\s_I]|}{n!} \frac{|[\s_J]|}{n!}\sum_R
    \frac{1}{f_R} \chi_R(\s_I) \chi_R(\s_J)
    \label{unnullcorrelator}
  \end{equation}
  This matrix provides the change of basis from the trace basis to the
  dual basis
  \begin{equation}
    \sum_J \corr{  \xi(\s_I, \Phi^\dagger)\xi(\s_J, \Phi) }  \tr(\s_J
    \Phi) = \xi(\s_I, \Phi^\dagger)
  \end{equation}
  where we sum $\sum_J$ over conjugacy classes of $S_n$. We have used
  identity \eqref{eq:sumsigmaorthogonal} of Section \ref{notation}.
  It follows that the matrix of correlators of the dual basis
  \eqref{unnullcorrelator} is the inverse of the matrix of correlators
  of the trace basis \eqref{untracecorrelator}
  \begin{equation}
    \sum_J \corr{ \xi(\s_I, \Phi^\dagger)\xi(\s_J, \Phi)
     }  \corr{  \tr(\s_J \Phi^\dagger)\tr(\s_K \Phi)
     }  =  \corr{  \xi(\s_I, \Phi^\dagger)\tr(\s_K \Phi)
     }   = \delta_{IK}
  \end{equation}

  In the large $N$ limit we see from equation \eqref{fR} that $f_R\to
  N^n$ so that the dual basis becomes, up to a factor, the trace basis
  \begin{equation}
    \xi(\s_I,\Phi)= \frac{|[\s_I]|}{n!} \sum_{R(n)}
    \frac{1}{f_R}\chi_R(\s_I) \chi_R(\Phi) \to \frac{|[\s_I]|}{N^nn!} \tr(\s_I\Phi)
  \end{equation}
  In this limit the duality of the two bases in equation
  \eqref{UNbasisisdual} is just the well-known orthogonality of traces
  at large $N$.

\end{enumerate}

\section{$SU(N)$ summary}\label{sunsum}

In $SU(N)$ our complex scalar is traceless.  Denote the $SU(N)$
complex scalar by $\Psi$ to distinguish it from the $U(N)$ complex
scalar $\Phi$ which does have a trace.  The correlator for $\Psi$ is
\begin{equation}
  \langle\;  \Psi^{\dagger}{}^i_{j} \Psi^k_l   \;\rangle  =
  \delta^i_l \delta^k_j - \frac{1}{N}\delta^i_j \delta^k_l
  \label{suninnerproduct}
\end{equation}
We can relate this to the $U(N)$ correlator \eqref{uninnerproduct} by
making the substitution $\Psi^i_j = \Phi^i_j - \delta^i_j \Phi^k_k/N$.
If we feed this substitution into the $U(N)$ correlator we get the
same result
\begin{align}
    \langle\;  \Psi^{\dagger}{}^i_{j} \Psi^k_l   \;\rangle = \langle\;
  \left( \Phi^{\dagger}{}^i_{j} - \delta^i_j \Phi^{\dagger}{}^m_{m}/N\right)\left( \Phi^k_{l} - \delta^k_l\Phi^n_{n}/N\right)  \;\rangle =
  \delta^i_l \delta^k_j - \frac{1}{N}\delta^i_j \delta^k_l
\end{align}
This means that we can use \emph{the same} correlator for both $U(N)$
and $SU(N)$, using operators built from $\Phi^i_j$ for $U(N)$ and from
$\Psi^i_j = \Phi^i_j - \delta^i_j \Phi^k_k/N$ for $SU(N)$.  This
ability to move between the $SU(N)$ and $U(N)$ correlators using the
substitution $\Psi^i_j = \Phi^i_j - \delta^i_j \Phi^k_k/N$ will be
extremely useful in later formulae. In essence this subsitution
enforces the tracelessness condition.\footnote{Note that this method
can also be applied to $O(N)$ and $Sp(2N)$.  Elements of the Lie
algebra of $O(N)$ are antisymmetric real matrices $\chi = -\chi^T$.
We can obtain the $O(N)$ correlator by the subsitution $\chi = i(X -
X^T)$ where $X$ is a hermitian generator of $U(N)$
(cf. \cite{Aharony:2002nd}).  Similarly for $Sp(2N)$ the real Lie
algebra elements $\Pi$ satisfy $J\Pi = (J\Pi)^T$ and their correlator
can be found with $\Pi = J(X + X^T)$.}

$\Psi$ is traceless $\tr \Psi = 0$ so we are going to need to consider
elements of $S_n$ without 1-cycles.  Define $C_n$ to be the subset of
$S_n$ with all the elements with 1-cycles removed.  For example
\begin{itemize}
\item $C_1 = \emptyset$
\item $C_2 = \{ (12) \}$
\item $C_3 = \{(123), (132)  \}$
\item $C_4 = \{[(12)(34)], [(1234)] \}$
\item $C_5 = \{[(12)(345)], [(12345)] \}$
\end{itemize}
$[(12)(34)]$ means the conjugacy class of $(12)(34)$, which is $\{
(12)(34), (13)(24), (14)(23)\}$.

Define a set of elements $\{\tau_i\}$ in $C_n$ where each $\tau_i$ is
an element of a different conjugacy class.  There are $p(n) - p(n-1)$
conjugacy classes in $C_n$, since each element with a 1-cycle can be
decomposed into a 1-cycle and an element of $S_{n-1}$.

The three bases of dimension $n$ gauge-invariant polynomials of $\Psi$
have some different properties to their $U(N)$ counterparts.
\begin{enumerate}
\item The \textbf{trace basis} is defined by the $p(n)-p(n-1)$
  conjugacy classes of $C_n$
  \begin{equation}
    \tr (\tau_i \Psi)
  \end{equation}
  For $n=2$ we have $\tr(\Psi^2)$, for $n=3$ we have $\tr(\Psi^3)$,
  for $n=4$ we have $\tr(\Psi^2)\tr(\Psi^2)$ and $\tr(\Psi^4)$ and for
  $n=5$ we have $\tr(\Psi^2)\tr(\Psi^3)$ and $\tr(\Psi^5)$.

\item The $p(n)$ elements of the \textbf{Schur polynomial basis}
  $\chi_R(\Psi)$ are now neither independent nor diagonal.  For each
  of the $p(n-1)$ Young diagrams $T$ with $n-1$ boxes we have a linear
  relation between the Schur polynomials of dimension $n$
  \begin{equation}
    0 = \tr(\Psi)\chi_T(\Psi)= \chi_{\yng(1)}(\Psi)\chi_T(\Psi) = \sum_{R(n)} g(\yng(1),T;R)
    \chi_R(\Psi)
  \end{equation}
  $\yng(1)$ is the single box representation $\chi_{\yng(1)}(\Psi) =
  \tr(\Psi) = 0$ and $g(\yng(1),T;R)$ is the Littlewood-Richardson
  coefficient for compositions of representations.  It is only
  non-zero if $R$ is in $\yng(1) \otimes T$.

\item The \textbf{dual basis} is defined by the $p(n)-p(n-1)$
  conjugacy classes of $C_n$
  \begin{equation}
    \xi(\t_i,\Psi):= \frac{|[\t_i]|}{n!} \sum_{R(n)}
    \frac{1}{f_R}\chi_R(\t_i) \chi_R(\Psi)
  \label{SUNnulldef}
  \end{equation}
  It turns out that even for $SU(N)$ this basis is dual to the trace
  basis using the inner product defined in \eqref{suninnerproduct},
  i.e.
  \begin{equation}
    \corr{  \xi (\t_i,\Psi^\dagger)  \tr (\t_j \Psi)  }  =
    \delta_{ij}
    \label{eq:dualtracenullSUN}
  \end{equation}
  We will show that for $SU(N)$ this dual basis is exactly the null
  basis constructed algorithmically in \cite{deMelloKoch}.

  The correlation function of two elements of the dual basis is given
  by
  \begin{equation}
    \corr{  \xi(\t_i,\Psi^\dagger)\xi(\t_j,\Psi) }  = \frac{|[\t_i]|}{n!}\frac{|[\t_j]|}{n!}\sum_R
    \frac{1}{f_R} \chi_R(\t_i) \chi_R(\t_j)
    \label{sunnullcorrelator}
  \end{equation}
  which is remarkably exactly the same as the $U(N)$ correlator of the
  dual basis \eqref{unnullcorrelator}, as proved in \cite{deMelloKoch}
  for the null basis.

  The matrix of correlators of the dual basis provides the change
  of basis from the trace basis to the dual
  \begin{equation}
    \sum_j \corr{  \xi(\t_i, \Psi^\dagger)\xi(\t_j, \Psi) }  \tr(\t_j
    \Psi) = \xi(\t_i, \Psi^\dagger)
    \label{sunchangebasis}
  \end{equation} 
  where we sum $\sum_j$ over conjugacy classes of $C_n$.  To get this
  result we can use the same argument as for the $U(N)$ case because
  we can add into the sum the remaining elements of $S_n$ with
  1-cycles, whose corresponding traces vanish.  Thus the matrix of
  correlators of the dual basis is also the inverse of the matrix of
  correlators of the trace basis
  \begin{equation}
    \sum_j \corr{ \xi(\t_i,\Psi^\dagger)\xi(\t_j,\Psi)
     }  \corr{  \tr(\t_j \Psi^\dagger)\tr(\t_k \Psi)
     }  = \corr{  \xi(\t_i, \Psi^\dagger)\tr(\t_k \Psi)
     }  = \delta_{ik}
  \end{equation}
  where again we sum $\sum_j$ over conjugacy classes of $C_n$.
\end{enumerate}

\section{$SU(N)$ details}\label{sundetails}

Following \cite{deMelloKoch} we define a derivative on the Schur
polynomials of a general $N \times N$ matrix $M^i_j$ by
\begin{equation}
  D \chi_R(M) = \sum_{i=1}^M \frac{\d}{\d M^i_i} \chi_R(M) = \sum_{T(n-1)}  g(\yng(1),T;R) \frac{f_R}{f_T}\chi_T(M)
\end{equation}
where we have given an exact formula for the derivative.  We sum over
representations $T$ with $(n-1)$ boxes that differ from $R$ by a
`legal' box.  $\yng(1)$ is the single-box fundamental representation;
$g(\yng(1),T;R)$ is a Littlewood-Richardson coefficient that is zero
if $R$ is not in $\yng(1) \otimes T$.  The formula for the
Littlewood-Richardson coefficient is given in Section \ref{notation}.
$\frac{f_R}{f_T}$ is the weight $(N-i+j)$ of the box removed from the
Young diagram of $R$ to get $T$, where $i$ labels the row and $j$
the column of the box in the Young diagram of $R$.

Using this we can Taylor expand for a constant $k$
\begin{align}
  \chi_R\left(M + k\mathbb{I} \right) & = \sum_{F=0}^n
   \frac{1}{F!}
  k^F  D^F \chi_R(M) \label{taylor}  \\
 & = \sum_{F=0}^n
   \frac{1}{F!}
  \sum_{T(n-f)}  g(\yng(1)^F,T;R) \frac{f_R}{f_T}k^F
  \chi_T(M)
\end{align}
Here $g(\yng(1)^F,T;R) = g(\yng(1), \dots \yng(1),T;R)$ with $F$
$\yng(1)$'s.  It counts the different legal ways we can build the
representation $R$ by adding $F$ single-box representations $\yng(1)$
to $T$.  $T$ has $(n-F)$ boxes.  For example
\begin{equation}
  g\left(\yng(1)^2,\yng(2)\; ;\yng(3,1)\right) = 2
\end{equation}
because
\begin{equation}
  \yng(1)\otimes \yng(1) \otimes \yng(2) = \yng(1)\otimes (\yng(3)
  \oplus \yng(2,1)) = 2 \; \yng(3,1) \oplus \yng(2,1,1) \oplus \yng(4)
  \oplus \yng(2,2)
\end{equation}

We have therefore
\begin{equation}
  \chi_R(\Psi) =  \chi_R\left(\Phi - \frac{\tr \Phi}{N} \mathbb{I} \right) = \sum_{F=0}^n
   \frac{1}{F!}
  \sum_{T(n-F)}  g(\yng(1)^F,T;R) \frac{f_R}{f_T} \left(-\frac{\tr\Phi}{N}\right)^{F}
  \chi_T(\Phi)
  \label{schurdecouplingSU}
\end{equation}
and conversely
\begin{equation}
  \chi_R(\Phi) =  \chi_R\left(\Psi + \frac{\tr \Phi}{N} \mathbb{I} \right) = \sum_{F=0}^n   \frac{1}{F!}
  \sum_{T(n-F)}  g(\yng(1)^F,T;R) \frac{f_R}{f_T}\left( \frac{\tr\Phi}{N}\right)^{F}
  \chi_T(\Psi)
  \label{schurdecouplingU}
\end{equation}
These two equations are entirely compatible.  If we feed the
expression for $\chi_T(\Psi)$ given by \eqref{schurdecouplingSU} into
\eqref{schurdecouplingU} we recover $\chi_R(\Phi)$.

In \cite{deMelloKoch} the authors algorithmically constructed a set of
operators annihilated by the operator $D$ which they called the null
basis.  Because they are annihilated by the operator $D$ the Taylor
expansion \eqref{taylor} is truncated to the $F=0$ terms.

Now we will show that that the $SU(N)$ dual basis $\xi(\tau_i,\Psi)$ for
$\tau_i \in C_n$ given in \eqref{SUNnulldef} is indeed null and hence,
using the substitution $\Psi = \Phi - \tr\Phi/N$, we have
\begin{equation}
  \xi(\tau_i,\Psi) = \xi(\tau_i,\Phi)
  \label{eq:SUNUNnullthesame}
\end{equation}
This is true because we get only the $F=0$ terms in the Taylor
expansion.

If we expand $\xi(\tau_i,\Psi)$
\begin{align}
  D \xi(\t_i,\Psi) & = \frac{|[\t_i]|}{n!} \sum_{R(n)}
  \frac{1}{f_R}\chi_R(\t_i)\, D \chi_R(\Psi) \nn \\
  & = \frac{|[\t_i]|}{n!}  \sum_{R(n)}\chi_R(\t_i)  \sum_{T(n-1)}  g(\yng(1),T;R) \frac{1}{f_T} \left(-\frac{\tr\Phi}{N}\right)
  \chi_T(\Phi)
  \label{nullexpansion}
\end{align}
This looks monstrous but if we extract the sum over $R$ and use the
identity \eqref{gRST} for $g(\yng(1),T;R)$ from Section
\ref{notation}, expanding it in characters of the symmetric group, we
see that
\begin{align}
  \sum_{R(n)}\chi_R(\t_i)g(\yng(1),T;R) & = \sum_{R(n)}\chi_R(\t_i) \frac{1}{(n-1)!} \sum_{\r\in S_{n-1}} \chi_{\yng(1)}(\id) \chi_T(\rho) \chi_R(\id \circ \r)\nn \\
& =  \frac{1}{(n-1)!} \sum_{\r\in
      S_{n-1}} \chi_{\yng(1)}(\id) \chi_T(\rho) \frac{n!}{|[\tau_i]|}
    \delta([\tau_i] = [\id \circ \r])
\end{align}
where we have used identity \eqref{eq:sumRorthogonal2}.  Here id is
the identity permutation made only of 1-cycles.  But we know that
$\t_i$ has no 1-cycles so $[\tau_i] = [\id\circ \r]$ is never
satisfied.  Therefore the $SU(N)$ dual basis is indeed null
$D\xi(\t_i, \Psi) = 0$ and thus $\xi(\tau_i,\Psi) = \xi(\tau_i,\Phi)$ is
true.

Note that this only works for the $SU(N)$ dual basis
$\xi(\tau_i,\Psi)$ with $\tau_i \in C_n$.  For a general $\s_I \in
S_n$ with 1-cycles, $\s_I \notin C_n$, $\xi(\s_I,\Psi)$ is not null
and we do not have $\xi(\s_I,\Psi) = \xi(\s_I,\Phi)$.

The correlator of two members of the $SU(N)$ dual basis
\eqref{sunnullcorrelator} now follows very quickly because it must be
the same as the $U(N)$ correlator
\begin{equation}
  \corr{  \xi(\t_i,\Psi^\dagger)\xi(\t_j,\Psi) }  = \corr{ 
  \xi(\t_i,\Phi^\dagger)\xi(\t_j,\Phi) }  = \frac{|[\t_i]|}{n!}\frac{|[\t_j]|}{n!}\sum_R
  \frac{1}{f_R} \chi_R(\t_i) \chi_R(\t_j)
\end{equation}

Using
\begin{equation}
  \corr{\Psi^\dagger \tr\Phi} = 0 \quad \Rightarrow\quad
  \corr{\Psi^\dagger \Psi} =  \corr{\Psi^\dagger \Phi}
  \label{unsunidentity}
\end{equation}
we can also see that the duality of the multi-trace
basis to the null basis follows from the $U(N)$ case
\begin{equation}
  \corr{  \xi (\t_i,\Psi^\dagger)  \tr (\t_j \Psi)  }  = \corr{ 
  \xi (\t_i,\Psi^\dagger)  \tr (\t_j \Phi)  } = \corr{  \xi
  (\t_i,\Phi^\dagger)  \tr (\t_j \Phi)  }  = 
  \delta_{ij}
\end{equation}
In the first equality we have used property \eqref{unsunidentity} that
$\corr{\Psi^\dagger \Psi} = \corr{\Psi^\dagger \Phi}$; in the second
we have used property \eqref{eq:SUNUNnullthesame} that
$\xi(\tau_i,\Psi) = \xi(\tau_i,\Phi)$; in the final inequality we have
used the defining property of the $U(N)$ dual basis
\eqref{UNbasisisdual}.

We would now like to show that the Schur polynomial basis is no longer
diagonal for $SU(N)$.  We can use \eqref{schurdecouplingU} to see that
\begin{eqnarray}
   &&\corr{\chi_R(\Phi^\dagger)\chi_S(\Phi)} \nn \\
&  = & \sum_{F=0}^n
   \frac{1}{(F!)^2} \frac{1}{N^{2F}}
  \sum_{T(n-F)} \sum_{U(n-F)} g(\yng(1)^F,T;R)  g(\yng(1)^F,U;S) \nn
  \\
& &  \times \frac{f_Rf_S}{f_Tf_U} \corr{\chi_{\yng(1)}^F(\Phi^\dagger)\chi_{\yng(1)}^F(\Phi)}\corr{
  \chi_T(\Psi^\dagger)\chi_U(\Psi)}\nn \\
&  = & \sum_{F=0}^n
   \frac{1}{(F!)}\frac{1}{N^F}
  \sum_{T(n-F)} \sum_{U(n-F)} g(\yng(1)^F,T;R)  g(\yng(1)^F,U;S)\frac{f_Rf_S}{f_Tf_U}\corr{
  \chi_T(\Psi^\dagger)\chi_U(\Psi)}
\end{eqnarray}

Separating out the $F=0$ term and re-arranging we see that
\begin{align}
  \corr{\chi_R(\Psi^\dagger)\chi_S(\Psi)} 
& = 
   \corr{\chi_R(\Phi^\dagger)\chi_S(\Phi)} \nn \\
&  - \sum_{F=1}^n
   \frac{1}{(F!)} \frac{1}{N^{F}}
  \sum_{T,U} g(\yng(1)^F,T;R)  g(\yng(1)^F,U;S)  \frac{f_Rf_S}{f_Tf_U}\corr{
  \chi_T(\Psi^\dagger)\chi_U(\Psi)}
\end{align}
which when applied recursively gives us
\begin{align}
   &\corr{\chi_R(\Psi^\dagger)\chi_S(\Psi)} \nn\\
& =  \sum_{F=0}^n
   \frac{1}{(F!)}\left( - \frac{1}{N}\right)^{F}
  \sum_{T,U} g(\yng(1)^F,T;R)  g(\yng(1)^F,U;S)\frac{f_Rf_S}{f_Tf_U}\corr{
  \chi_T(\Phi^\dagger)\chi_U(\Phi)} \nn \\
& = 
 \sum_{F=0}^n
   \frac{1}{(F!)}\left( - \frac{1}{N}\right)^{F}
  \sum_{T}  g(\yng(1)^F,T;R)  g(\yng(1)^F,T;S)\frac{f_Rf_S}{f_T}
\end{align}
This agrees with the calculation in equation (10.7) of \cite{cr} if we
make the identification $g(\yng(1)^F,T;R) = \sum_U d_U g(U,T;R)$.
This identification follows from the identities in Section
\ref{notation} and the fact that $d_U = \chi_U(\id^{\circ F})$.
The formula also agrees with the results from \cite{deMelloKoch}.

\section{Factorisation and probabilities for $SU(N)$}\label{factor}

Given that we have a basis and its dual we can write down
factorisation equations for $SU(N)$ correlators analogous to those
described in $\cite{copto}$ for $U(N)$ correlators.  For a conformal
field theory like $\cN=4$ super Yang-Mills these factorisation
equations let us write correlators on 4-dimensional surfaces with
non-trivial topology in terms of correlators on the 4-sphere, just
like factorisation of correlators on Riemann surfaces in two
dimensions.  Because of positivity properties of the summands in the
factorisation equations we can interpret these summands as
well-defined probabilities for a large class of processes.  Since we
are only interested in the combinatorics we will drop the spacetime
dependences and any extraneous modular parameters.

If a complete basis for the local operators of our $SU(N)$ theory is
given by $\{ \cO_a \}$ and the metric on this basis from the two point
function has an inverse $G^{ab}$, then for local operators $\cA$,
$\cB$ the sphere factorisation is given by a sum of positive
quantities $\cite{copto}$
\begin{align}
  \corr{\cA^\dagger\cB}  & = \sum_{a,b} G^{ab} 
  \corr{\cA^\dagger  \cO_a}\big\langle \;\cO_b^\dagger \cB \;\big\rangle  \nn\\
 & > \sum_{i,j} \cG^{ij} 
  \corr{\cA^\dagger  \tr(\t_i \Psi)}\corr{ \tr(\t_j \Psi^\dagger) \cB} \nn\\
& = \sum_{i}
  \corr{\cA^\dagger \tr(\t_i \Psi)}\corr{ \xi(\t_i,\Psi^\dagger)
  \cB}
\label{giraffe}
\end{align}
We have truncated the sum over operators of the $SU(N)$ theory to
those half-BPS operators made from a single complex scalar $\Psi$.  In
the sum $i$ and $j$ range over the conjugacy classes of $C_n$.  We
have used the fact that the inverse of the metric on the trace basis
is the correlator of the dual basis $ \cG^{ij} = \corr{
  \xi(\t_i,\Psi^\dagger) \xi(\t_j,\Psi) } $, which effects the change
of basis from the trace basis to the dual basis
\eqref{sunchangebasis}.  If we set $\cB = \cA$ and divide both sides
of \eqref{giraffe} by $\corr{\cA^\dagger \cA}$ we get a sum of
well-defined, positive probabilities
\begin{align}
  P\left(\cA \to \tr(\t_i \Psi)\right) & = P\left(\cA \to
  \xi(\t_i, \Psi)\right) =  \frac{\corr{\cA^\dagger \tr(\t_i \Psi)}\corr{ \xi(\t_i,\Psi^\dagger) \cA}}{\corr{\cA^\dagger \cA}}
\end{align}

If one of $\cA$ and $\cB$ is a polynomial in $\Psi$ then we can
connect the $SU(N)$ factorisation \eqref{giraffe} to the $U(N)$
factorisation.  The first step is to use $\xi(\t_i,\Psi) =
\xi(\t_i,\Phi)$
\begin{equation}
  \sum_{i} \corr{\cA(\Psi^\dagger) \tr(\t_i \Psi)}\corr{ \xi(\t_i,
 \Psi^\dagger) \cB } =  \sum_{i} \corr{\cA(\Psi^\dagger) \tr(\t_i \Psi)}\corr{ \xi(\t_i,
 \Phi^\dagger) \cB }
\end{equation}
Because $\tr(\Psi) = 0$ we can add back in the
conjugacy classes of $S_n$ with 1-cycles since these terms are zero
\begin{equation}
 \sum_{i} \corr{\cA(\Psi^\dagger) \tr(\t_i \Psi)}\corr{ \xi(\t_i,
 \Phi^\dagger) \cB } =  \sum_{I} \corr{\cA(\Psi^\dagger) \tr(\s_I \Psi)}\corr{ \xi(\s_I,
 \Phi^\dagger) \cB }
\end{equation}
Here $I$ ranges over the conjugacy classes of $S_n$. Finally we use
$\corr{\Psi^\dagger \Psi} = \corr{\Psi^\dagger \Phi}$ to see that
$\corr{\cA(\Psi^\dagger) \tr(\t_i \Psi)} = \corr{\cA(\Psi^\dagger)
\tr(\t_i \Phi)}$ and hence
\begin{equation}
\sum_{I} \corr{\cA(\Psi^\dagger) \tr(\s_I \Psi)}\corr{ \xi(\s_I,
 \Phi^\dagger) \cB } = \sum_{I} \corr{\cA(\Psi^\dagger) \tr(\s_I \Phi)}\corr{ \xi(\s_I,
 \Phi^\dagger) \cB }
\label{unfactfromsun}
\end{equation}
This is now a sum over $U(N)$ operators, which gives us the $U(N)$
factorisation.  This only works if one of $\cA$ and $\cB$ is a
function of $\Psi$.  If $\s_I$ contains 1-cycles the summand vanishes
because $\corr{\Psi^\dagger \tr(\Phi)} = 0$.  So what we are really
saying is that if one of $\cA$ and $\cB$ is a polynomial in $\Psi =
\Phi - \tr(\Phi)/N$ we can truncate the $U(N)$ factorisation
\eqref{unfactfromsun} to the $SU(N)$ factorisation \eqref{giraffe}.
If we translate this into probabilities it means that
$P\left(\cA(\Psi) \to \tr(\t_i \Psi)\right) =
P\left(\cA(\Psi) \to \tr(\t_i \Phi)\right)$.

Since $\xi(\t_j,\Psi) = \xi(\t_j,\Phi)$ is a polynomial in $\Psi$ we
find the probability
\begin{equation}
  P\left(\xi(\t_j,\Psi) \to \tr(\t_i \Psi)\right)  = \delta_{ij}
\end{equation}
which is exactly the same as the corresponding $U(N)$ result
$P\left(\xi(\t_j,\Phi) \to \tr(\t_i \Phi)\right)$.

For a transition into two separate states we use the factorisation on
a 4-dimensonal `genus one' surface
\begin{align}
  \corr{\cA^\dagger \cB}_{G=1} & > \sum_{i,j} \sum_{k,l} \cG^{ij} \cG^{kl}
  \corr{\cA^\dagger \tr(\t_i \Psi) \tr(\t_k \Psi)}\corr{ \tr(\t_l
  \Psi^\dagger) \tr(\t_j \Psi^\dagger) \cB} \nn\\
& = \sum_{i} \sum_{k}
  \corr{\cA^\dagger \tr(\t_i \Psi) \tr(\t_k \Psi)}\corr{ \xi(\t_k,\Psi^\dagger) \xi(\t_i,\Psi^\dagger) \cB}
\end{align}
If one of $\cA$ and $\cB$ is a function of $\Psi$ then the $U(N)$
factorisation truncates to this result.

The probability of a transition to KK gravitons is given by
\begin{align}
  P\left(\cA \to \tr(\t_i \Psi),\tr(\t_k \Psi)\right)   =   \frac{\corr{\cA^\dagger \tr(\t_i \Psi) \tr(\t_k \Psi)}\corr{
      \xi(\t_k,\Psi^\dagger) \xi(\t_i,\Psi^\dagger)
      \cA}}{\corr{\cA^\dagger \cA}_{G=1}}
\end{align}

For $\cA = \xi(\t_m,\Psi)$
\begin{align}
&  P\left(\xi(\t_m,\Psi) \to \tr(\t_i \Psi),\tr(\t_k
  \Psi)\right)\\
& =   \frac{\corr{\xi(\t_m,\Psi^\dagger) \tr(\t_i \Psi) \tr(\t_k \Psi)}\corr{ \xi(\t_k,
  \Psi^\dagger) \xi(\t_i \Psi^\dagger)     
      \xi(\t_m,\Psi)}}{\corr{\xi(\t_m,\Psi^\dagger)
      \xi(\t_m,\Psi)}_{G=1}} \nn \\
& =   \frac{\delta_{[\t_m] = [\t_i
    \circ \t_k]}\corr{ \xi(\t_k,
  \Psi^\dagger) \xi(\t_i \Psi^\dagger)     
      \xi(\t_m,\Psi)}}{\corr{\xi(\t_m,\Psi^\dagger)
      \xi(\t_m,\Psi)}_{G=1}}
\end{align}
So $\cA = \xi(\t_m,\Psi)$ will decay into two multi-trace operators as
long as $\t_i \circ \t_k$ is in the conjugacy class of $\t_m$.

\section{Diagonalisation by higher Hamiltonians}\label{diag}

In this section we will find a diagonal basis for the $SU(N)$
correlator.\footnote{This section was done in collaboration with
Sanjaye Ramgoolam.}  We can reduce the half-BPS sector of the $\cN =
4$ SYM to matrix quantum mechanics \cite{cjr,Berenstein:2004kk}. For
gauge group $U(N)$ the Schur polynomials are eigenstates of commuting
higher Hamiltonians (for $U(N)$ these correspond to the Casimirs of
the Lie algebra).  Our strategy will be to find eigenstates of the
higher Hamiltonians for $SU(N)$.  These eigenstates are necessarily
diagonal.

If we do a reduction of the $\cN = 4$ SYM action on $S^3$ with only
the first two real scalars $X_1$ and $X_2$ turned on then we get a
(0+1)-dimensional matrix model
\begin{equation}
  \label{eq:reducedaction}
  S = \int dt  \Tr( \dot X_1^2 +  \dot X_2^2 - X_1^2
  -X_2^2 ).
\end{equation}
The potential term couples to the curvature of $S^3$ but we have
rescaled the fields appropriately.  If we introduce the complex chiral
scalar $Z= X_1+ i X_2$ and find its momentum conjugate $\Pi$ then we
can define harmonic oscillator operators $A = Z + i\Pi$ and $B = Z -
i\Pi$ and their conjugates $A^\dagger$ and $B^\dagger$.  These satisfy
standard commutation relations
\begin{equation}
  [A_a, A^\dagger_b] =  g_{ab}
\end{equation}
where $a,b$ run over the adjoint representation of the gauge group and
$g_{ab}$ is the inverse of the bilinear invariant form $g^{ab} = \tr(T^a
T^b)$.

Our Hamiltonian is
\begin{equation}
  \label{eq:hamiltonianmatrix}
  H = \tr(A^\dagger A + B^\dagger B)
\end{equation}
and our angular momentum operator is
\begin{equation}
  \label{eq:angmommatrix}
  J = \tr(A^\dagger A- B^\dagger B)
\end{equation}

For $\tr((A^\dagger)^n(B^\dagger)^m)\vac$, $E= n+m$, $J = n-m$.  For
our highest weight chiral primaries we have $E= J$ so $m$ is zero and
we restrict to the $\tr((A^\dagger)^n)\vac$ states.  We have higher
Hamiltonians
\begin{equation}
  H_n = \tr((A^\dagger A)^n)  
\end{equation}
that commute with $H= \tr(A^\dagger A)$.

If we concentrate on the $U(N)$ case we find that in terms of adjoint
matrix indices
\begin{equation}
  \label{eq:UNliealgebra}
  [A^i_j, A^\dagger{}^k_l] = [A_a,A^\dagger_b](T^a)^i_j(T^b)^k_l =
  g_{ab}(T^a)^i_j(T^b)^k_l =  \delta^i_l \delta^k_j
\end{equation}

The Schur polynomials are simultaneous eigenstates of these higher
Hamiltonians and the different eigenvalues give a complete
identification of each Schur polynomial
\begin{equation}
  \label{eq:evaluehigherH}
  H_n \chi_R(A^\dagger) \vac = C_n^R \chi_R(A^\dagger) \vac
\end{equation}
For $U(N)$ these higher Hamiltonians are in fact the Casimirs of the
Lie algebra (cf. \cite{Barut:1986dd}).

We can show that these Schur polynomials are diagonal.  Suppose we
make no assumptions about the correlator of the Schur polynomials and
call it $h_{RS}:=\bra{0}\chi_R(A)\chi_S(A^\dagger)\vac$.
\begin{equation}
  \label{eq:hamiltoniansandwiche}
  \bra{0}\chi_R(A)H_n \chi_S(A^\dagger)\vac = C^S_n h_{RS} = C^R_n h_{RS}
\end{equation}
We have acted to the right with $H_n$ and then to the left.  If
$h_{RS}\neq 0$ then we must have $C^R_n = C^S_n$ for all $n$;
otherwise $h_{RS}= 0$.  We have enough Casimirs to distinguish between
the Schur polynomials so if $R \neq S$ then $C^R_n \neq C^S_n$ for
some $n$, so we must have $h_{RS}= 0$ for $R \neq S$.

Now extend this argument to the $SU(N)$ case for which
\begin{equation}
  \label{eq:SUNliealgebra}
[A^i_j, A^\dagger{}^k_l]  = g_{ab}(T^a)^i_j(T^b)^k_l = \delta^i_l \delta^k_j -
\frac{1}{N}\delta^i_j \delta^k_l
\end{equation}

The higher Hamiltonians no longer have simple eigenvectors or
eigenvalues.  Also the higher Hamiltonians no longer correspond to the
Casimirs of $SU(N)$.  However they must diagonalise the correlator by
the same argument as above.

For example, at level 4 we have two independent gauge-invariant states
for which
\begin{align}
 H \tr(A^{\dagger 2})\tr(A^{\dagger 2})\vac & = 4\tr(A^{\dagger
   2})\tr(A^{\dagger 2})\vac  \nn \\
 H \tr(A^{\dagger 4})\vac  & = 4 \tr(A^{\dagger 4})\vac  \nn \\
 H_2 \tr(A^{\dagger 2})\tr(A^{\dagger 2})\vac  & = \left[\left(4N -
  \frac{8}{N}\right)\tr(A^{\dagger 2})\tr(A^{\dagger 2}) + 8\tr(A^{\dagger 4})\right]\vac \nn \\  
H_2 \tr(A^{\dagger 4})\vac  & = \left[\left(4+\frac{12}{N^2}\right)\tr(A^{\dagger 2})\tr(A^{\dagger 2}) + \left(4N
  -\frac{28}{N}\right)\tr(A^{\dagger 4})\right]\vac
\end{align}
If we find the eigenvectors of $H_2$, we get a diagonal basis
\begin{align}
  \label{eq:diagSUNbasis}
  \left[\left(\frac{5}{4N}-\frac{\sqrt{49N^2+8N^4}}{4N^2}\right)\tr(A^{\dagger 2})\tr(A^{\dagger 2}) +\tr(A^{\dagger 4})\right]\vac \\
  \left[\left(\frac{5}{4N}+\frac{\sqrt{49N^2+8N^4}}{4N^2}\right)\tr(A^{\dagger 2})\tr(A^{\dagger 2}) +\tr(A^{\dagger 4})\right]\vac.
\end{align}

This method of using eigenvectors of higher Hamiltonians to
diagonalise the correlator will work at all levels.  While it is as
complicated as a Gram-Schmidt diagonalisation, it does at least share
its derivation from the higher Hamiltonians with the $U(N)$ matrix
model.

\vskip.5in

{ \bf Acknowledgements } I would like to thank Sanjaye Ramgoolam for
suggesting this project and for his many helpful suggestions.  In
addition I am grateful to Robert de Mello Koch, Se\'an Murray and Nick
Toumbas for useful discussions.

\begin{appendix}

\section{Useful identities}\label{notation}

There are two orthogonality relations for the characters of the
symmetric group $S_n$.
\begin{enumerate}
\item For $\rho,\tau\in S_n$ if we sum over representations $R$ of $S_n$
  \begin{equation}
    \label{eq:sumRorthogonal2}
    \sum_{R(n)} \chi_R (\rho) \chi_R (\tau) =\frac{n!}{|[\tau]|}
    \delta([\tau] = [\rho])
  \end{equation}
  This generalises for $\rho_i \in S_{n_i}$ and $\tau \in S_n$ where
  $n = \sum_i n_i$ and $i=1, 2 \dots k$
  \begin{align}
    \label{eq:sumRorthogonal}
    \sum_{R_1(n_1), R_2(n_2) \dots R_k(n_k), T(n)} g(R_1, R_2, \dots
    R_k;T) \chi_{R_1}(\rho_1) \chi_{R_2}(\rho_2) \cdots
    \chi_{R_k}(\rho_k)\chi_{T}(\tau)\nn\\ 
     = \frac{n!}{|[\tau]|}
    \delta([\t] = [\rho_1\circ \rho_2 \cdots \circ \rho_k])
  \end{align}

\item For representations $R,S$ of $S_n$ if we sum over the elements
  of $S_n$
  \begin{equation}
    \label{eq:sumsigmaorthogonal}
    \sum_{\s\in S_n} \chi_R (\s) \chi_S (\s) =  n! \delta_{RS}
  \end{equation}
  This generalises for representations $R_i$ of $S_{n_i}$
  \begin{align}
    \label{gRST}
     \frac{1}{n_{R_1}!\cdots n_{R_k}!} \sum_{\r_1\in
      S_{n_1}}\cdots \sum_{\r_k\in S_{n_k}} \chi_{R_1}(\r_1) \cdots
    \chi_{R_k}(\rho_k) \chi_T(\r_1 \circ \cdots \r_k) \nn \\
     =  g(R_1,\dots R_k;T)
  \end{align}
  which is the Littlewood-Richardson coefficient.
\end{enumerate}
We can also derive from \eqref{gRST}
\begin{equation}
  \chi_T(\rho_1 \circ \cdots \circ \rho_k) = \sum_{R_1(n_1)} \cdots
  \sum_{R_k(n_k)} g(R_1, \cdots R_k; T) \chi_{R_1}(\rho_1) \cdots \chi_{R_k}(\rho_k)
\end{equation}

\end{appendix}

\end{document}